\documentclass[aps,prd,preprint,nofootinbib,showpacs,amssymb]{revtex4}
\usepackage{amsmath,amsfonts}

\newcommand{\be}{\begin{equation}}
\newcommand{\ee}{\end{equation}}
\newcommand{\bea}{\begin{eqnarray}}
\newcommand{\eea}{\end{eqnarray}}

\begin{document}

\author{Gerhard Sch\"afer}
\email{G.Schaefer@tpi.uni-jena.de}

\author{ Achamveedu Gopakumar}
\email{A.Gopakumar@tpi.uni-jena.de}

\affiliation{Theoretisch-Physikalisches Institut,
Friedrich-Schiller-Universit\"at, Max-Wien-Platz 1, 07743 Jena, Germany}

\title{A minimal no-radiation approximation to Einstein's field equations}

\date{\today}

\begin{abstract}
An approximation to Einstein's field equations
in  Arnowitt-Deser-Misner (ADM) canonical formalism is presented which
corresponds to the magneto-hydrodynamics (MHD) approximation
in electrodynamics.
It results in coupled elliptic equations which represent
the maximum of elliptic-type structure of Einstein's theory
and naturally generalizes previous conformal-flat truncations of the
theory. The Hamiltonian, in this approximation, is identical with
the non-dissipative part of the Einsteinian one through the third post-Newtonian order.
The proposed scheme, where stationary spacetimes are  exactly reproduced,
should be useful to construct {\em realistic} initial data for 
general relativistic simulations as well as to
model astrophysical scenarios, where gravitational radiation reaction
can be neglected.
%perform
%relativistic simulations which neglect gravitational
%radiation.

\end{abstract}

\pacs{04.20.Ex,04.20.Cv,04.25.Dm,04.30.Db}

\maketitle

\section{Introduction}

In electrodynamics, to model 
quasi-stationary scenarios accurately,
it is customary to drop
the Maxwell displacement current in  Maxwell-Amp\`ere law
of Maxwell's equations.
This approximation is often referred to as
the magneto-hydrodynamics (MHD) approximation
as it is mainly implemented in  MHD computations 
dealing, mostly, with astrophysical plasmas.
A strictly
analogous approximation to 
Einstein's general relativity,
should prove very useful to model 
non-stationary situations, involving 
relativistic self gravitating systems.
%for  Einstein's general relativity.

%In electrodynamics,  it is well known that dropping
%the Maxwell displacement current in  Maxwell-Amp\`ere equation,
%often simply refered to as the magneto-hydrodynamics (MHD) approximation
%because of its most prominent implementation there,
%is an excellent scheme to represent quasi-stationary situations.
%Therefore, a strictly
%analogous approximation should prove very useful
%for  Einstein's general relativity.

In Einstein's theory of gravity,  the conformal-flat approximations          
played a major role within numerical relativity as they resulted in
%relatively 
simple elliptic equations \cite{C00}.
However, in the past, it was pointed out that 
these approximations  are rather crude
for highly non-spherical objects like rotating disks of dust or
compact binary systems \cite{KS00,R96}. Therefore, in the following,
we shall propose an elliptic-type approximation
which goes far beyond conformal-flat schemes. Whereas
the conformal-flat approximations are not exact under stationarity
conditions, in the new approach,
Einstein stationarity is reproduced exactly. In a post-Newtonian
setting, the new approach, in general, does not reproduce all
even-order post-Newtonian terms (of course, the odd terms are not part
of the new approach) beyond the second post-Newtonian order. However,
at the third post-Newtonian order of approximation,
it still gives the 
equations of motion, derivable from the full Einstein theory \cite{S85,JS98}.
%(cf. \cite{S85}, \cite{JS98}). 
%The crucial equation of the present paper is the Eq. (16), or (18).
We observe that the very recently
proposed gravito-anelastic
approximation is quite different
from our minimal no-radiation  approximation, especially
for the choice of the 
dropped and the kept time derivatives in the 
evolution equations \cite{BGGN}. Our approximation also results 
in elliptic equations, 
which should be solvable employing LORENE,
a highly efficient numerical library, based on spectral methods,
developed by the Meudon group \cite{Lorene}.
% Eric Gourgoulhon et. al. \cite{Lorene}. 
%is of opposite type. 

\section{ Einstein's theory in the ADM formalism}
In the ADM formulation of general relativity, the spacetime line element
in the $(3+1)$ decomposed form is given by
%Let the metric of the spacetime be denoted by ($i,j = 1,2,3; ct=x^0$)
\be
ds^2 = - \alpha^2 c^2 dt^2 + \gamma_{ij}(dx^i + \beta^icdt)(dx^j+\beta^jcdt),
\mbox{ ( $i,j = 1,2,3 $ ) },
\ee
where $\alpha$ is the lapse function, $\beta^i$ the shift vector,
$\gamma_{ij}$ the induced metric on a three-dimensional 
spatial slice $\Sigma (t)$, parametrized by the
time coordinate $t$, and $c$ is the velocity of
light. The 3-metric and its canonical conjugate 
$ \frac {c^3 }{ 16\,\pi\,G}\,\pi^{ij}$, which is 
a contravariant symmetric tensor
density of weight $+1$, satisfy  the Hamiltonian and  momentum constraints
\cite{ADM62}
\be
\gamma^{1/2} \mbox{R} = \frac{1}{2\gamma^{1/2}}(2\pi^i_j \pi^j_i-\pi^2)
+\frac{16\pi G}{c^4} \gamma^{1/2} \alpha^2 T^{00}\,,
\label{Eq.2.2}
\ee
\be
-2 \pi^j_{i,j} + \pi^{kl} \gamma_{kl,i} = \frac{16\pi G}{c^4}
\gamma^{1/2}  \alpha T^0_i\,,
\label{Eq.2.3}
\ee
where  $\mbox{R}$ is 
the curvature scalar of $\Sigma (t)$, 
$\gamma$ the determinant of $\gamma_{ij}$,
$\pi^i_j = \gamma_{jk}\pi^{ik}$, $\pi = \pi^i_i$. 
$T^{00}$ and $ T^0_i$ are the components of 
the unspecified  4-dimensional stress-energy tensor for the matter, 
$T^{\mu\nu}$.
%and $G$ the Newtonian gravitational constant.
The canonical conjugate $\pi^{ij} $ is related to $K_{ij}$,  the extrinsic 
curvature of $\Sigma(t)$, by 
$\pi^{ij} = - \gamma^{1/2}
(\gamma^{il} \gamma^{jm} - \gamma^{ij} \gamma^{lm}) K_{lm}$,
where $ \gamma^{il} $ is the inverse metric of $\gamma_{ij}$.
In the above equations, 
a partial derivative is denoted by $,$
and  $G$ is the Newtonian gravitational constant.
We note that the more familiar form for the left hand side of 
Eq. (\ref{Eq.2.3}), namely  $ \pi^j_{i|j}$, where $|$ 
stands for the  3-dimensional covariant derivative, is expanded as
$\pi^j_{i,j} - \frac{1}{2} \pi^{kl} \gamma_{kl,i} $,
using the fact that $\pi^j_i$ is a mixed tensor density of weight $+1$.
The 3-metric and its canonical conjugate evolve in accordance with
the following evolution equations
\cite{ADM62}
\bea
\pi^{ij}_{~~,0} &=& -\frac{1}{2}\alpha \gamma^{1/2} (2\mbox{R}^{ij} - \gamma^{ij}\mbox{R}) 
+ \frac{1}{4}\alpha\gamma^{-1/2}\gamma^{ij}(2\pi^m_n \pi^n_m-\pi^2)
%\nonumber\\[1ex] &-& 
-\alpha \gamma^{-1/2}(2\pi^{im} \pi^j_m-\pi\pi^{ij})
\nonumber \\
&&
+  \gamma^{1/2}(\alpha^{|ij}-\gamma^{ij}\alpha^{|m}_{~~~|m})
%\nonumber\\[1ex] &+& 
+ (\pi^{ij}\beta^m)_{|m} - \pi^{mj} \beta^i_{~|m} - \pi^{mi} \beta^j_{~|m}
\nonumber \\
&&
+ \frac{8\pi G}{c^4} ~ \alpha \gamma^{1/2} \gamma^{il}\gamma^{jm} T_{lm}
\label{Eq.2.4}
\eea
and 
\be
\gamma_{ij,0} = \alpha \gamma^{-1/2}(2\pi_{ij} - \gamma_{ij}\pi) + \beta_{i|j} +
\beta_{j|i},
\label{Eq.2.5}
\ee
where $\mbox{R}_{ij}$ is the Ricci
tensor associated with $\Sigma (t)$.
%and the $|$ stands for the  3-dimensional covariant derivative.
In this paper, we raise and lower indices
on 3-dimensional objects with $\gamma^{ij}$ and  $\gamma_{ij}$
respectively.

     The ADM coordinate conditions which generalize the isotropic
Schwarzschild metric read
\be
\pi^{ii}=0,
\label{Eq.2.6}
\ee
where, and from here onwards, repeated covariant or contravariant
indices imply the usage of Einstein summation convention and
\be
\gamma_{ij}= \psi^4\,\delta_{ij}+h^{\rm TT}_{ ij},
\label{Eq.2.7}
\ee
where $\psi$ is a conformal scalar and $h^{TT}_{ij}$ 
the transverse-traceless (TT) part of the
3-metric $\gamma_{ij}$ with respect to the euclidean 3-metric
$\delta_{ij}$. By definition, $h^{TT}_{ij}$ satisfies 
%following relations,
$h^{\rm TT}_{ii} = h^{\rm TT}_{ ij,j} = 0$.
The restriction $h^{\rm TT}_{ij} = 0$ gives 
a simple expression for the 3-dimensional curvature scalar,
$\gamma^{1/2} \mbox{R} = - 8\,\psi\,\Delta \psi$,
where $\Delta$ stands for the 3-dimensional euclidean Laplacian.
The differential equation, used for gauge fixing $\gamma_{ij}$,
reads
\be
3 \gamma_{ij,j} - \gamma_{jj,i} = 0. 
\label{Eq.2.8}
\ee
Taking into account the gauge condition for $\pi^{ij}$, namely
Eq.~(\ref{Eq.2.6}), 
the following decomposition can be achieved,
\be
\pi^{ij}={\tilde{\pi}^{ij}}+\pi_{\rm TT}^{ij},
\label{Eq.2.9}
\ee
where ${\tilde{\pi}^{ij}}$ denotes the longitudinal part of $\pi^{ij}$. It may 
be expressed as   
\be
\tilde{\pi}^{ij}= \pi^j_{~,i} + \pi^i_{~,j} - \frac{2}{3}\delta_{ij} \pi^m_{~~,m}, 
\label{Eq.2.10}
\ee
which implies $\pi^{ij}_{~~,j} \equiv \tilde{\pi}^{ij}_{~~,j} 
= \Delta \pi^i + \frac{1}{3}\pi^j_{i,j}$, suggesting that 
Eq. (\ref{Eq.2.3}) can be used to compute an  elliptic equation for
$\pi^i$.
%i.e. $\pi^{ij}_{~~,j} = \tilde{\pi}^{ij}_{~~,j} i
%= \Delta \pi^i + (1/3)\pi^j_{j,i}$.
The TT-part $\pi^{ij}$, namely 
$ \frac{ c^3 }{ 16\,\pi\,G} \pi_{\rm TT}^{ij}$,
is the canonical conjugate to $h^{\rm TT}_{ij}$, which gives 
the independent field degrees of freedom. 
The Hamiltonian, which  generates the time
evolution of the independent degrees of freedom of the system
(matter plus gravitational field), is given by \cite{ADM62}
%The Hamilton functional of the total system is given by
\be
\mbox{H}[MV, h^{\rm TT}_{ij}, \pi_{\rm TT}^{ij}]
= - \frac{c^4}{2\pi G}\int d^3x~
\Delta\psi [MV, h^{\rm TT}_{ij}, \pi_{\rm TT}^{ij}],
\label{Eq.2.11}
\ee
where $MV$ denotes the (non-specified) matter variables. 
If the matter system consists of black holes, the matter 
variables $MV$ in our paper enter via boundary conditions at 
the apparent horizons and at spacelike infinity.

The Eqs. (\ref{Eq.2.6}) and (\ref{Eq.2.7}) 
result in the covariant trace of $\pi^{ij}$ of the
form $\pi =\pi^{ij} h^{\rm TT}_{ij}$. Taking into account the
space-asymptotic properties  $\pi^{ij} \sim 1/r^2$ and $h^{\rm TT}_{ij} \sim
1/r$, the gauge condition Eq. (6) turns out to mean asymptotic maximal
slicing. The gauge conditions Eqs. (6) and (7), or (8), are very close to
the well-known Dirac gauge conditions. The condition Eq. (8), e.g., is
identical with the corresponding Dirac gauge condition to linear order
in $\gamma_{ij} - \delta_{ij}$.

The functions $\psi$ and $\pi^i$,
and hence $\tilde{\pi}^{ij}$, are
determined using the Hamiltonian and momentum constraints,
given by  Eqs. (\ref{Eq.2.2}) and (\ref{Eq.2.3}),  by elliptic equations. 
The elliptic equations for the (scalar) lapse $\alpha$  and the 
(vector) shift functions 
$\beta^i$  result from the Eqs. (\ref{Eq.2.4}) and (\ref{Eq.2.5}), 
applying the coordinate conditions Eqs. (\ref{Eq.2.6}) and (\ref{Eq.2.7}) 
respectively.
%(8),
The explicit Poisson-type equations for $\alpha$
and $\beta^i$ can be derived from 
\bea
\gamma^{ii}\alpha^{|m}_{~~~|m} - \alpha^{|ii} =
&-& \alpha (\mbox{R}^{ii} - \gamma^{ii}\mbox{R})
- 2 \alpha \gamma^{-1}\pi^{im} \pi^i_m 
- 2 \gamma^{-1/2} \pi^{im} \beta^i_{~|m} 
%\nonumber\\[1ex] &+&
\nonumber \\
&&
+ \frac{8\pi G}{c^4} ~ \alpha (\gamma^{il}\gamma^{im} T_{lm}-
\gamma^{ii} \alpha^2T^{00}),
\label{Eq.2.12}
\eea
where use has been made of Eq. (\ref{Eq.2.2}).
The elliptic equation for $\beta^i$ results from
\bea
(\beta_{i|j})_{,j} + (\beta_{j|i})_{,j} - \frac{2}{3} (\beta_{j|j})_{,i}
& = & - (\alpha \gamma^{-1/2} (2\pi_{ij} - \gamma_{ij}\pi))_{,j} 
%\nonumber\\[1ex] &+& 
+ \frac{1}{3}(\alpha \gamma^{-1/2} (2\pi_{jj} - \gamma_{jj}\pi))_{,i}. 
\label{Eq.2.13}
\eea
The functions, 
$h^{\rm TT}_{ij}$ and $\pi_{\rm TT}^{ij}$, 
in general,  follow from the evolution equations, given by
Eqs. (\ref{Eq.2.4}) and (\ref{Eq.2.5}).
%(5) in the form of evolutionary equations.
Alternatively,  $h^{\rm
TT}_{ij}$ and $\pi_{\rm TT}^{ij}$ may be obtained from the Hamiltonian,
Eq. (\ref{Eq.2.11}), via the corresponding Hamilton equations. 
 
\section{The minimal no-radiation approximation}

Motivated by the MHD  approximation in electrodynamics, a
minimal truncation of the Einstein theory, in view of suppressing
radiation emission, is achieved by putting {\em only} the TT-part for the 
evolution equation for $\pi^{ij}$, given by 
Eq. (\ref{Eq.2.4}),  to zero. 
The TT-part of Eq. (\ref{Eq.2.4})  may be denoted as 
\be
\pi^{ij}_{\rm TT,0} = A^{ij}_{TT}\,,
\label{Eq.3.1}
\ee
where $A^{ij}_{TT}  \equiv  A^{ij} - A^{ij}_L$.
The expression for $A^{ij}$,  obtainable from the right hand side  
of Eq. (\ref{Eq.2.4}), after using the Hamiltonian constraint equation, 
%given by Eq. (2),
reads
\bea
A^{ij} \equiv &-& \alpha \gamma^{1/2} (\mbox{R}^{ij} - \gamma^{ij} \mbox{R})
- \alpha \gamma^{-1/2}(2\pi^{im} \pi^j_m-\pi\pi^{ij})
%\nonumber\\[1ex] &+& 
+ \gamma^{1/2}(\alpha^{|ij}-\gamma^{ij}\alpha^{|m}_{~~~|m})
\nonumber \\
&+& (\pi^{ij}\beta^m)_{|m} - \pi^{mj} \beta^i_{~|m} - \pi^{mi} \beta^j_{~|m}
%\nonumber\\[1ex] &+& 
+ \frac{8\pi G}{c^4} ~ \alpha \gamma^{1/2}(\gamma^{il}\gamma^{jm} T_{lm}
- \gamma^{ij} \alpha^2 T^{00}).
\label{Eq.3.2}
\eea
The longitudinal part of  $A^{ij}$, namely $A^{ij}_L$,
may be written in terms of a vector $ V^i$ as
\be
A^{ij}_L = V^j_{~,i} + V^i_{~,j} - \frac{2}{3}\delta_{ij} V^m_{~~,m} 
\label{Eq.3.3}
\ee
which satisfies
\be
\Delta V^i + \frac{1}{3} V^j_{~~,ji} =   A^{ij}_{~,j}\,.
\label{Eq.3.4}
\ee
The elliptic equation that  determines $h^{\rm TT}_{ij}$, 
derivable  using the condition $ A^{ij}_{TT} = 0 $, reads
%and it reads,
%which determines the waveless $h^{\rm TT}_{ij}$ reads
%(minimal waveless condition) 
\bea
2\gamma \mbox{R}_{ij} = &-&  \frac{2\gamma^{1/2}}{\alpha} \gamma_{il}
\gamma_{jn} \biggl [A^{ln}_L
- (\pi^{ln}\beta^m)_{|m} + \pi^{lm} \beta^n_{|m} + \pi^{nm} \beta^l_{|m}
\biggr ]
% \nonumber\\[1ex] &+&
+ 
\gamma_{ij}(2\pi^m_n \pi^n_m-\pi^2)
\nonumber \\
&&
 - 4\pi^{m}_i \pi_{jm} + 2\,\pi\pi_{ij} 
%\nonumber \\ &+&
+ \frac{2\gamma}{\alpha} (\alpha_{|ij}-\gamma_{ij}\alpha^{|m}_{~~~|m}) 
% \nonumber\ \[1ex] &+& 
+ \frac{16\pi G}{c^4} \gamma (T_{ij} + \gamma_{ij} \alpha^2 T^{00})\,.
\label{Eq.3.5}
\eea
The  left hand side of the above equation, which introduces the Laplacian 
for $h^{\rm TT}_{ij}$, takes the form 
\bea
2 \gamma \mbox{R}_{ij}  &=&
- \chi^8 \biggl [
\Delta h^{\rm TT}_{ij} +  \Delta (\psi^4) \delta_{ij} + (\psi^4)_{,ij} \biggr]
%\nonumber\\[1ex] &+&
+ (\psi^4 h^{\rm TT}_{kl}  -   h^{\rm TT}_{mk} h^{\rm TT}_{ml})
\biggl ( 
\gamma_{kl,ij}  + \gamma_{ij,kl} 
\nonumber \\
&&
- \gamma_{kj,il} - \gamma_{il,kj} \biggr )
%\nonumber\\[2ex] &+&
+ 2 \gamma \gamma^{kl} \gamma^{np}
(\Gamma_{n.il} \Gamma_{p.kj} - \Gamma_{n.ij} \Gamma_{p.kl})\,,
\label{Eq.3.a}
\eea 
where $ \chi^8 = \psi^8 - \frac{1}{2} h^{\rm TT}_{mn}h^{\rm TT}_{mn} $
and where the Christoffel symbols $\Gamma_{i.jk}$ are given by
$ \frac{1}{2} \, ( \gamma_{ij,k}  
+ \gamma_{ik,j} - \gamma_{jk,i} )$. For the derivation of 
Eq. (\ref{Eq.3.a}), we have also used the following
relation $\gamma \gamma^{ij} = \chi^8 \delta_{ij}
- \psi^4 h^{\rm TT}_{ij}  + 
h^{\rm TT}_{ik} h^{\rm TT}_{kj} $ \cite{S85}.
In terms of a function $\phi$, vanishing at spacelike infinity,
$\psi$ may be given by $ (1 + \frac{\phi}{8})$.

%\begin{equation}
%2 \Gamma_{i.jk} = \gamma_{ij,k}  + \gamma_{ik,j} - \gamma_{jk,i},
%\end{equation}
%and where use has been made of 
%\begin{equation}
%\gamma \gamma^{ij} = \chi^8 \delta_{ij}
%-  \psi^4 h^{\rm TT}_{ij}  +   h^{\rm TT}_{ik} h^{\rm TT}_{kj}
%\end{equation}
%with 
%\begin{equation}
%\chi^8 = \psi^8 - \frac{1}{2} h^{\rm TT}_{mn}h^{\rm TT}_{mn}.
%\end{equation}
%In terms of a function $\phi$ vanishing at spacelike infinity, $\psi$
%may read,   
%\begin{equation}
%\psi = 1 + \frac{\phi}{8}.
%\end{equation}

In the Hamiltonian formulation,
the present approximation to  Einstein's equations is identical with the
replacement of only the evolution equation
\be
\pi^{ij}_{\rm TT,0} = - \frac{16 \pi G}{c^4} \frac{\delta
\mbox{H}}{\delta h^{\rm TT}_{ij}},
\label{Eq.3.6}
\ee
which is the TT-part of the Eq. (\ref{Eq.2.4}), through 
\be
\frac{\delta \mbox{H}}{\delta h^{\rm TT}_{ij}} = 0.
\label{Eq.3.7}
\ee
 
    The dropping of $\pi^{ij}_{\rm TT,0}$ exactly corresponds to the dropping
of the displacement current,
$ \frac{1 }{ c}\, \frac{\partial {\bf E} }{ \partial t}$,
in  Maxwell-Amp\`ere equation
of  electrodynamics. This is so, since $-{\bf E} $ is 
the momentum conjugate in the Hamiltonian formulation 
of electrodynamics. 
However, this is not quite the way MHD like approximation to 
general relativity is implemented in \cite{BGGN}, as they 
impose  $h^{\rm TT}_{ij,0} = 0$ in their scheme.
%Moreover, this will note generate, 
In the post-Newtonian approximation, their scheme will not 
generate the correct leading order expression for
$\pi^{ij}_{\rm TT}$. Moreover, the dynamics associated
with their approximation will coincide with 
the Einsteinian one through second post-Newtonian order
only.

%and at the 
%third post-Newtonain order, the associated dynamics would not
%even coincide with  the Einsteinian one.
%The way the analogy
%is defined in \cite{BGGN} we are not supporting. In following \cite{BGGN}
%we would have to put $h^{\rm TT}_{ij,0} = 0$ in Eq. (21).  But this
%would not yield to the correct leading-order post-Newtonian expression
%for $\pi^{ij}_{\rm TT}$, see \cite{JS98}; also its third post-Newtonian
%dynamics would not coincide with th gv waveless.ps
%e Einsteinian one.   
    
    It is obvious that stationary spacetimes are reproduced exactly
in our approximation.
The constraint and  the evolution equations, given by 
Eqs. (\ref{Eq.2.2}), (\ref{Eq.2.3}), (\ref{Eq.2.4}), and (\ref{Eq.2.5}) 
still determine
$\psi$, $\pi^i$, $\alpha$, and $\beta^i$ respectively 
%The field equations, Eqs. (2) and (3), 
%and the constraint equations, Eqs. (4) and (5),  still determine
%$\phi$ and $\pi^i$
through elliptic equations. Moreover,
Eq. (\ref{Eq.3.5}) defines $h^{\rm TT}_{ij}$, which describes the independent field degrees of freedom,
through an elliptic equation, (see Eq. (3.8a) in \cite{S85},
for more details).
In addition, 
the longitudinal part of $  \pi^{ij}_{~~,0}$,
defined via  $V^i$, 
can be determined through an elliptic equation for 
$V^i$, given by Eq. (17).
The remaining quantities $\pi^{ij}_{\rm TT}$
should be determined, solely algebraically,
using the evolution equation for $\gamma_{ij}$, given by Eq. (\ref{Eq.2.5}),
in the following manner.
%the Eq. (16) determines
%$h^{\rm TT}_{ij}$ through an elliptic equation too (cf. eq. (3.8a) in
%\cite{S85}) and $\pi^{ij}_{\rm TT}$ is determined through the Eq. (3)
%solely algebraically.  Writing  
Let us write the Eq. (5) in the form
\be
\gamma_{ij,0} = B_{ij}\,,
\label{Eq.3.8}
\ee
where 
\be
B_{ij} \equiv  \alpha \gamma^{-1/2}(2\pi_{ij} - \gamma_{ij}\pi) + \beta_{i|j}
+ \beta_{j|i}.
\ee
Using the definition of $\gamma_{ij}$, given by Eq. (7), 
we obtain,    
\be
h^{\rm TT}_{ij,0}  =  B_{ij} - \frac{1}{3}B_{ll}\delta_{ij}
\ee
%Hereof, by the aid of the equations 
where 
$ B_{ll} = 3 (\gamma^{mn}\gamma^{mn})^{-1}(B^{ll} 
- \gamma^{il} \gamma^{ik} h^{\rm TT}_{lk,0}) $ with  $B^{ll}$ given by
$ - \alpha \gamma ^{-1/2} \gamma^{ll} \pi + 2 \beta^{l|l} $.
%\be
%B_{ll} = 3 (\gamma^{mn}\gamma^{mn})^{-1}(B^{ll} - \gamma^{il} \gamma^{ik} h^{\rm TT}_{lk,0})
%\ee
%and
%\be
%B^{ll} = - \alpha \gamma ^{-1/2} \gamma^{ll} \pi + 2 \beta^{l|l},
%\ee
Using these equations,     
$\pi^{ij}_{\rm TT}$ follows in terms of 
$h^{\rm TT}_{ij,0}$, the only time derivative appearing 
in the minimal no-radiation
approximation, given by
\bea
\pi^{ij}_{\rm TT} = &-& \tilde{\pi}^{ij}  
%\nonumber\\[1ex] &-&
-\frac{\gamma^{1/2}}{\alpha} C^{ijlk} [\beta^{l|k} - 
\beta^{m|m}  \gamma^{ln} \gamma^{kn} (\gamma^{pq} \gamma^{pq})^{-1}] \nonumber\\[1ex]
&+&  \frac{\gamma^{1/2}}{2\alpha}  C^{ijlk} [\gamma^{lm} \gamma^{kn} -
 \gamma^{lr} \gamma^{kr} \gamma^{sm} \gamma^{sn}
(\gamma^{pq} \gamma^{pq})^{-1}]h^{\rm TT}_{mn,0},
\label{Eq.3.11}
\eea
where $C^{ijlk}$ is the inverse matrix of the matrix   
\be
C^{-1}_{ijlk} \equiv  \frac{1}{2} \left(\delta_{il}\delta_{jk}   + \delta_{ik}\delta_{jl} -  
(\gamma^{ij} - \gamma^{in} \gamma^{jn}  \gamma^{mm} (\gamma^{pq} \gamma^{pq})^{-1})
h^{\rm TT}_{lk} \right).
\ee
Notice, the matrix $C^{-1}_{ijlk}$ deviates from the unit matrix in the
quadratic order of $h^{\rm TT}_{lk}$ only.
%For maximal slicing
%condition, $0 = \gamma^{ij}\,K_{ij} = \pi \gamma^{-1/2}/2$, 
%the matrix  $C^{-1}_{ijlk}$ would be the unit matrix.
For maximal slicing condition, $0 = \gamma^{ij}\,K_{ij} 
= \pi \gamma^{-1/2}/2$, 
%and 
the resulting 
$C^{-1}_{ijlk}$  would be a unit matrix,
when applied to symmetric tensors.

The internal consistency of the proposed approach comes from the fact that
only the evolution equations for the true field degrees of freedom are
affected. This  implies, in the evolution equations for the TT variables,
%and this can % done only uniquely
%without spoiling their internal consistency %i.e. 
\begin{subequations}
\begin{align}
\pi^{ij}_{\rm TT,0} =& - \frac{16 \pi G}{c^4} \frac{\delta
\mbox{H}}{\delta h^{\rm TT}_{ij}}, 
\label{Eq.3.15a}
\\
%\quad \quad 
h_{ij,0}^{\rm TT} =&~\, \frac{c^4}{16 \pi G}\frac{\delta \mbox{H}}{\delta \pi_{\rm TT}^{ij}},
\label{Eq.3.15b}
\end{align}
\end{subequations}
%without spoiling their internal consistency, {\em i.e.} 
if the first equation is solved for  $h^{\rm TT}_{ij}$ 
using Eq. (\ref{Eq.3.5}), then the second equation 
can be solved for $\pi_{\rm TT}^{ij}$
in terms of $ \dot{h}_{ij}^{\rm TT} $, as given by
Eq. (\ref{Eq.3.11}).
However, putting $\dot{h}_{ij}^{\rm TT} = 0 $ in Eq. (\ref{Eq.3.15b}) 
means destroying the Legendre transformation, which is
a fundamental property of the physical theory, 
whereas setting $\dot{\pi}^{ij}_{\rm TT} = 0 $ in Eq. (\ref{Eq.3.15a})
implies a dynamical condition for the extremum (minimum) of energy.
Imposing $h^{\rm TT}_{ij} =0$, instead of solving for it, results
in the well-known Wilson-Mathews approach, if  $\pi^{ij}_{\rm TT}$ is
determined through Eq. (\ref{Eq.3.8}). 
We also note that a recent effort, the so-called CFC+ approach,
which tries to improve  conformal-flat approximations
uses Eqs (\ref{Eq.3.5}) and (\ref{Eq.3.11}),
restricted to the leading post-Newtonian order, 
to get 
$h^{\rm TT}_{ij}$ and $\pi^{ij}_{\rm TT}$ \cite{TBCD03}.
This approach is identical with
the full Einstein theory through the second post-Newtonian order.
%The leading order improved
%Conformal Flat Condition approach, the CFC+ approach,
%now under development by numerical researchers,  
%e.g. see \cite{TBCD03},  is obtained by solving the Eqs. (18) and (19)
%for $h^{\rm TT}_{ij}$ and  the Eq. (27) for $\pi^{ij}_{\rm TT}$ to
%leading post-Newtonian order. The CFC+ approach is identical with
%the full Einstein theory through second post-Newtonian order.
 
In general, the Hamiltonian for the matter system is the Routh
functional depending on the matter variables only \cite{JS98,S03},
\be
{\cal{R}}[MV, h^{\rm TT}_{ij}(MV), h^{\rm TT}_{ij,0}(MV)] =
\mbox{H} - \frac{c^3}{16\pi G}\int d^3x  \pi^{ij}_{\rm TT} h^{\rm TT}_{ij,0}.
\ee
In the stationary case, ${\cal{R}}$ and $ \mbox{H}$ coincide.
%If the matter system consists of black holes, the matter 
%variables $MV$ in our paper enter via boundary conditions at 
%the apparent horizons and at spacelike infinity.
For our quasi-stationary approximation, we also shall take  ${\cal{R}}$
as Hamiltonian. It agrees with the non-dissipative part of the
Einsteinian one through third post-Newtonian order as e.g. given
in Ref. \cite{JS98}. 

   Finally, we point out that  $\dot {h}_{ij}^{\rm TT}$ is the {\em only}
partial time derivative appearing in the coupled elliptic equations,
present in the minimal no-radiation approximation.
%As $\dot{h}_{ij}^{\rm TT}$ is a small quantity,
While employing any iterative procedure
to solve our coupled elliptic equations,
it should be natural to equate 
the (in general) small quantity  $ \dot{h}_{ij}^{\rm TT}$
to zero to obtain 
the first order solution to $ {h}_{ij}^{\rm TT}$ and the rest of 
the variables involved. 
During the second stage of iteration,
one should then employ $\dot{h}_{ij}^{\rm TT}$,
computed using the first order ${h}_{ij}^{\rm TT}$, to get
the second order solution to $ {h}_{ij}^{\rm TT}$ and other 
quantities.
In this manner, the iterative procedure may be extended to higher 
stages.  The above procedure was employed, 
in the post-Newtonian framework, first  to compute 
$h^{\rm TT}_{ij}$ to the second post-Newtonian order,
which was employed to compute $\dot{h}_{ij}^{\rm TT}$,
required at the third post-Newtonian order 
calculations \cite{JS98}.

%  Finally it should be pointed out that in all of our elliptic equations
%  there is involved the partial time derivative of $h^{\rm TT}_{ij}$
%  only. 
%As  $\dot{h}_{ij}^{\rm TT}$, in general, is quite a small quantity, it is
%  natural to put it to zero for a first-order solution of the elliptic
%  equations.  The plugging in of $\dot{h}_{ij}^{\rm TT}$ from the
%  first-order solution allows the calculation of a second-order one;
%  and so on. In the  paper Ref. \cite{JS98} this procedure was
% applied within a post-Newtonian framework: $h^{\rm TT}_{ij}$
%  was calculated at the second post-Newtonian order first, and
%  $\dot{h}_{ij}^{\rm TT}$ contributed then to the third post-Newtonian
%  one.

   This communication will be followed by a detailed article \cite{GS_03},
where the explicit expressions, in terms of the
basic variables and their spatial derivatives, for the elliptic equations for 
$\psi, \alpha, \beta^i$, and $ h^{\rm TT}_{ij}$ as well as  for $\pi^i$ and
$V^i$ will be presented.

\begin{acknowledgments}

We thank E. Gourgoulhon  and L. Rezzolla for useful remarks.
The work is supported by the Deutsche
Forschungsgemeinschaft (DFG) through SFB/TR7
``Gravitationswellenastronomie''. 

\end{acknowledgments}

\end{document}